\documentclass{sig-alternate-2013}

\permission{\copyright 2017 International World Wide Web Conference Committee \\ (IW3C2), published under Creative Commons CC BY 4.0 License.}
\conferenceinfo{WWW 2017,}{April 3--7, 2017, Perth, Australia.}
\copyrightetc{ACM \the\acmcopyr}
\crdata{978-1-4503-4913-0/17/04. \\
http://dx.doi.org/10.1145/3038912.3052634 \\
\includegraphics{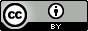}
}

\usepackage[pass,]{geometry}
\newif\iffull\fulltrue 
\newif\ifremarks\remarksfalse
\newif\ifdraft\draftfalse 

\usepackage[font=bf,skip=\baselineskip]{caption}
\usepackage[inline,shortlabels]{enumitem}

\usepackage[utf8]{inputenc}
\usepackage{color}
\definecolor{lightgray}{rgb}{0.95, 0.95, 0.95}
\definecolor{darkgray}{rgb}{0.4, 0.4, 0.4}
\definecolor{purple}{rgb}{0.65, 0.12, 0.82}
\definecolor{editorGray}{rgb}{0.95, 0.95, 0.95}
\definecolor{editorOcher}{rgb}{1, 0.5, 0} 
\definecolor{editorGreen}{rgb}{0, 0.5, 0} 
\usepackage{listings}
\usepackage{hyperref}
\usepackage{amsmath}
\usepackage{amssymb}
\usepackage{multirow}
\usepackage[usenames,dvipsnames,svgnames,table]{xcolor}

\lstdefinelanguage{CSS}{
  keywords={color,background-image:,margin,padding,font,weight,display,position,top,left,right,bottom,list,style,border,size,white,space,min,width, transition:, transform:, transition-property, transition-duration, transition-timing-function},	
  sensitive=true,
  morecomment=[l]{//},
  morecomment=[s]{/*}{*/},
  morestring=[b]',
  morestring=[b]",
  alsoletter={:},
  alsodigit={-}
}

\lstdefinelanguage{CSP}{
  keywords={default-src, script-src, child-src, frame-src, script-src, style-src, report-uri, connect-src, img-src, object-src, frame-ancestors, plugin-types, form-action, sandbox, worker-src, font-src, media-src},
  morestring=[b]',
  alsoletter={:},
  alsodigit={-}
}

\lstdefinelanguage{JavaScript}{
  morekeywords={typeof, new, true, false, catch, function, return, null, catch, switch, var, if, in, while, do, else, case, break},
  morecomment=[s]{/*}{*/},
  morecomment=[l]//,
  morestring=[b]",
  morestring=[b]'
}

\lstdefinelanguage{HTML5}{
  language=html,
  sensitive=true,	
  alsoletter={<>=-},	
  morecomment=[s]{<!-}{-->},
  tag=[s],
  otherkeywords={
  >,
	<!DOCTYPE,
  </html, <html, <head, <title, </title, <style, </style, <link, </head, <meta, />,
	</body, <body,
	</div, <div, </div>, 
	</p, <p, </p>,
	</script, <script,
  <canvas, /canvas>, <svg, <rect, <animateTransform, </rect>, </svg>, <video, <source, <iframe, </iframe>, </video>, <image, </image>
  },
  ndkeywords={
  =,
  charset=, src=, id=, width=, height=, style=, type=, rel=, href=,
  fill=, attributeName=, begin=, dur=, from=, to=, poster=, controls=, x=, y=, repeatCount=, xlink:href=,
  margin:, padding:, background-image:, border:, top:, left:, position:, width:, height:,
  transform:, -moz-transform:, -webkit-transform:,
  animation:, -webkit-animation:,
  transition:,  transition-duration:, transition-property:, transition-timing-function:,
  }
}

\def\scrsrc{\textbf{script-src} }
\def\con	src{\textbf{connect-src} }

\def\objsrc{\textbf{object-src} }

\def\frasrc{\textbf{frame-ancestors} }
\def\sansrc{\textbf{sandbox }}
\def\srcdoc{\textbf{srcdoc} }
\def\allowsop{\textbf{allow-same-origin} }
\def\allowsrc{\textbf{allow-scripts} }
\def\docdom{\code{document.domain}}

\def\none{\textbf{'none'} }
\def\self{\textbf{'self'} }
\def\star{\textbf{*} }
\def\uninline{\textbf{'unsafe-inline'} }
\def\uneval{\textbf{'unsafe-eval'} }

\def\code#1{\texttt {{#1}}}
\def\lib#1{\texttt {{#1}}}

\def\SHORTEN{\vspace*{-0.5cm}}

\def\CSP{\ensuremath{\mathcal{C}}}

\def\Frames{\ensuremath{\mathcal{F}}}
\def\HomeFrames{\ensuremath{\Frames}}
\def\Links{\ensuremath{L}}
\def\LinkedFrames{\ensuremath{\Frames_\Links}}
\def\CSPH{\ensuremath{\CSP}}
\def\CSPL{\ensuremath{\CSP_L}}
\def\CSPFH{\ensuremath{\CSP_\mathit{F}}}
\def\CSPFL{\ensuremath{\CSP_\mathit{LF}}}

\begin{document}

\title{On the Content Security Policy Violations due to the Same-Origin Policy}

\numberofauthors{3}
\author{
\alignauthor
Doli\`{e}re Francis Some\\
       \affaddr{Universit\'e C\^ote d'Azur}\\
       \affaddr{Inria, France}\\
       \email{doliere.some@inria.fr}
\alignauthor
Nataliia Bielova\\
       \affaddr{Universit\'e C\^ote d'Azur}\\
       \affaddr{Inria, France}\\
       \email{nataliia.bielova@inria.fr}
\alignauthor Tamara Rezk\\
       \affaddr{Universit\'e C\^ote d'Azur}\\
       \affaddr{Inria, France}\\
       \email{tamara.rezk@inria.fr}
}
  \maketitle
  
  \begin{abstract}
Modern browsers implement different security policies such as the Content Security Policy (CSP), a  mechanism designed to mitigate  
popular web vulnerabilities, and the Same Origin Policy (SOP), a mechanism that governs interactions between resources of web pages.

In this work, we describe how  CSP may be violated due to the SOP when a page contains an embedded  iframe from the same origin. 
We analyse 1 million pages from 10,000 top Alexa sites and report  that  at least 31.1\% of current CSP-enabled pages are potentially vulnerable to CSP violations.
Further considering real-world situations where those pages are involved in same-origin nested browsing contexts, we found that in at least 23.5\% of the cases, CSP violations are possible.

During our study, we also identified a divergence among browsers implementations in the enforcement of CSP in srcdoc sandboxed iframes, which actually reveals a problem in Gecko-based browsers CSP implementation.
To ameliorate the problematic conflicts of the security mechanisms, we discuss measures to avoid CSP violations.
\end{abstract}


  \label{sec:intro}
    \section{Introduction}
Modern browsers implement different  specifications to securely fetch and integrate content.
One widely used specification to protect content  is 
the Same Origin Policy (SOP)\cite{Same-Origin-Policy}. 
SOP allows developers to  isolate untrusted content from a different origin. 
An origin here is defined as scheme, host, and port number.  
If  an iframe's content  is loaded from a different origin,  SOP controls the access to the embedder resources.
In particular, no script inside the iframe can access content of the embedder page.
However, if the iframe's content is loaded from the same origin as the embedder page, there are no privilege restrictions w.r.t. the embedder resources. 
In such a case,  a script executing inside the iframe can access content of the embedder webpage. 
Scripts are considered trusted and the {\it iframe becomes transparent} from a developer view point.
A more recent specification to protect content in web pages is the Content Security Policy (CSP)\cite{Stam-Ster-Mark-10-WWW}.
The primary goal of CSP is to mitigate cross site scripting attacks (XSS), data leaks attacks, and other types of attacks.
CSP allows developers to specify, among other features, trusted domain sources from which to fetch content.
One of the most important features of CSP, is to allow a web application developer to specify  trusted  JavaScript sources. 
This kind of restriction  is meant to permit execution of  only trusted code and thus prevent untrusted code to access content of the page.

\begin{figure}[!t]
\includegraphics[width=0.45\textwidth, scale=0.3 ]{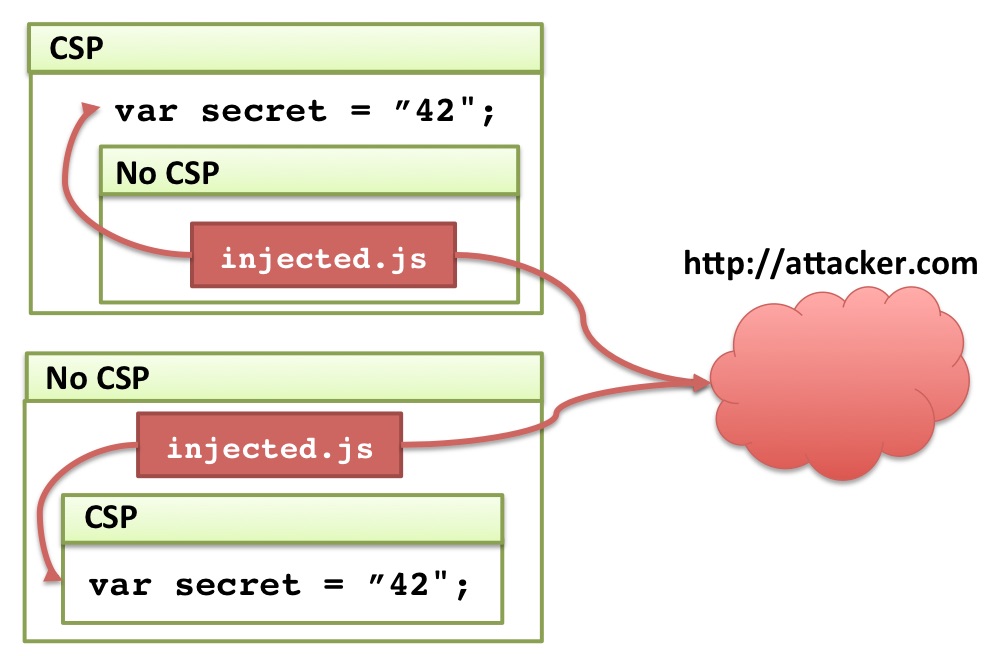}
\caption{An XSS attack
despite CSP.}
\label{fig:attack}
\SHORTEN
\end{figure}

In this work, we report on a fundamental problem of CSP. 
CSP\cite{CSP2-W3C} defines how to protect content in an isolated page. However, it does not  take into consideration the page's context, that is its embedder or embedded iframes.
In particular, CSP is unable to protect content of its corresponding page if the page embeds (using the {\it src} attribute) an iframe of the same origin. 
The CSP policy of a page will not be applied to an embedded iframe. 
However, due to SOP, the iframe has complete access to the content of its embedder. 
Because same origin iframes  are transparent due to SOP, 
this opens  loopholes to  attackers whenever the CSP policy of an iframe and that of its embedder page  are not  compatible (see Fig.~\ref{fig:attack}). 

We analysed 1 million pages from the top 10,000 Alexa sites and found that 5.29\% of sites contain some pages with CSPs (as opposed to 2\% of home pages in previous studies\cite{Calz-Rabi-Bugl-16-CCS}).
We have identified that in  94\% of cases, CSP may be violated in presence of the document.domain API and in 23.5\% of cases CSP may be violated without any assumptions (see Table~\ref{table:csp_violations}). 

During our study, we also identified a divergence among browsers implementations in the enforcement of CSP\cite{CSP2-W3C} in sandboxed iframes embedded with \emph{srcdoc}. 
This actually reveals an inconsistency between the CSP and HTML5 sandbox attribute specification for iframes.

We identify and discuss possible solutions  from the developer point of view as well as new security specifications that can help prevent this kind of CSP violations.
We have made publicly available the dataset that we used for our results in\cite{CSP_Violations}.
We have installed an automatic crawler to recover the same dataset every month to repeat the experiment taking into account the time variable. 
An accompanying technical report with a complete account of our analyses can be found at\cite{CSP-SOP-full}.

In summary, our contributions are: (i) We describe a new class of vulnerabilities  that lead to CSP violations. (Section~\ref{sec:cspsop}). 
(ii) We perform a large and depth scale crawl of top sites, highlighting CSP adoption at sites-level, as well as sites origins levels.
  Using this dataset,  we report on the possibilities of CSP violations  between the SOP and CSP in the wild. (Section~\ref{sec:study}). 
(iii) We propose guidelines in the design and deployment of CSP. (Section~\ref{sec:countermeasures}). 
  (iv) We reveal an inconsistency between the CSP specification and HTML5 sandbox attribute specification for iframes. Different browsers choose to follow different specifications, and we explain how any of these choices can lead to new vulnerabilities. (Section~\ref{sec:inconsistencies}).

  \label{sec:cspsop}
  \section{Content Security Policy and SOP}
The Content Security Policy (CSP)\cite{Stam-Ster-Mark-10-WWW} is a mechanism that allows programmers 
to control which client-side resources can be loaded and executed by the browser. 
CSP (version 2) is an official W3C candidate recommendation\cite{CSP2-W3C}, and is currently supported by  major web browsers. CSP is delivered in the \code{Content-Security-Policy} HTTP response header,  or 
in a \code{<meta>} element of HTML. 

{\bf CSP applicability}
A CSP  delivered with a page controls the resources of 
the page. However it does not apply to the page's embedding resources\cite{CSP2-W3C}. 
As such, CSP does not control the content of an iframe even if the 
iframe is from the same origin as the main page according to SOP. 
Instead, the content of the iframe is controlled by the CSP delivered 
with it, that can be different from the CSP of the main page. 

{\bf CSP directives}
CSP allows a programmer to specify which resources are 
allowed to be loaded and executed in the page. These resources are defined 
as a set of origins and known as a \emph{source list}. Additionally to controlling 
resources, CSP allows to specify allowed destinations of the AJAX requests by the 
\code{connect-src} directive. A special header \code{Content-Security-Policy-Report-Only} configures 
a CSP in a report-only mode: violations are recorded, but not enforced.
The directive \code{default-src} is a special fallback directive that is used when some directive is not defined. 
The directive \code{frame-ancestors} (meant to supplant the HTTP \code{X-Frame-Options} header\cite{CSP2-W3C}), controls in which pages the current page 
may be included as an iframe, to prevent clickjacking attacks\cite{framebust}.
See Table~\ref{tab:directives} for the most commonly used CSP directives\cite{Weic-etal-16-CCS}.

\begin{table}[!ht]
	\begin{tabular}{p{2.5cm}|p{5.5cm}}
	{\bf Directive} & {\bf Controlled content} \\
	\hline
	\code{script-src} & Scripts \\
	\code{default-src} & All resources (fallback) \\	
	\code{style-src} & Stylesheets \\
	\code{img-src} & Images \\
	\code{font-src} & Fonts \\
	\code{connect-src} & XMLHttpRequest, WebSocket \ or EventSource\\
	\code{object-src} & Plug-in formats (object, embed) \\	
	\code{report-uri} & URL where to report CSP violations \\
	\code{media-src} & Media (audio, video) \\
	\code{child-src} & Documents (frames), [Shared] Workers \\
	\code{frame-ancestors} & Embedding context \\
	\end{tabular}
\caption{Most common CSP directives\cite{Weic-etal-16-CCS}.}
\label{tab:directives}
\SHORTEN
\end{table}

{\bf Source lists}
CSP source list is traditionally defined as a \emph{whitelist} indicating 
which domains are trusted to load the content, or to communicate. 
For example, a CSP from Listing~\ref{lst:CSPex} allows to include scripts 
only from \lib{third.com}, requires to load frames only over HTTPS, while 
other resource types can only be loaded from the same hosting domain.  

\begin{lstlisting}[caption={Example of a CSP policy.}, label={lst:CSPex}]
Content-Security-Policy: default-src 'self'; 
script-src third.com; child-src https:
\end{lstlisting}

A whitelist can be composed of concrete hostnames (\lib{third.com}), 
may include a wildcard \code{*} to extend the policy to subdomains (\lib{*.third.com}), 
a special keyword \code{'self'} for the same hosting domain, 
or \code{'none'} to prohibit any resource loading.

{\bf Restrictions on scripts}
Directive \code{script-src} is the most used feature of CSP in today's web applications\cite{Weic-etal-16-CCS}. 
It allows a programmer to control the origin of scripts in his 
application using source lists. 
%
When the \code{script-src} directive is present in CSP, it blocks the execution of any 
inline script, JavaScript event handlers and APIs that execute string data code, such as 
\code{eval()} and other related APIs. 
To relax the CSP, by allowing the execution of inline \code{<script>} and JavaScript event handlers, 
a \code{script-src} whitelist should contain a keyword \code{'unsafe-inline'}.  
To allow \code{eval()}-like APIs, the CSP should contain a \code{'unsafe-eval'} keyword.
Because \code{'unsafe-inline'} allows execution of \emph{any} inlined script, it effectively 
removes any protection against XSS. Therefore, nonces and hashes were introduced in 
CSP version 2\cite{CSP2-W3C}, 
allowing to control which inline scripts can be loaded and executed. 

{\bf Sandboxing iframes}
Directive \code{sandbox} allows to load resources but execute them in a separate  
environment. It applies to all the iframes and other content present on the page. 
An empty \code{sandbox} value creates completely isolated iframes. One can selectively enable specific features via
\code{allow-*} flags in the directive's value. For example, \code{allow-scripts} 
will allow executions of scripts in an iframe, and \code{allow-same-origin} will allow iframes to be treated as being from their normal origins. 

\paragraph{Same-Site and Same-Origin Definitions}
In our terminology, we distinguish the web pages that 
belong to the same site from the pages that belong to the same origin. 
By \emph{page} we refer to any HTML document -- for example, 
the content of an iframe we call \emph{iframe page}. In this case, 
the page that embeds an iframe is called a \emph{parent page} 
or \emph{embedder}. 

By \emph{site} we refer to the highest level domain that we extract from 
Alexa top 10,000 sites, usually containing the domain name and a TLD, 
for example \lib{main.com}. All the pages that belong to a site, and to any of its subdomains as \lib{sub.main.com}, 
 are considered \emph{same-site} pages.

According to the Same Origin Policy, an \emph{origin} of a page is scheme, host 
and port of its URL. For example, in \url{http://main.com:81/dir/p.html}, the scheme is ``http'', the host is ``main.com'' 
and the port is 81.

\subsection{CSP violations due to SOP}
\label{sec:example}

   Consider a web application, where the main page \lib{A.html} and its iframe \lib{B.html} 
   are located at \lib{http://main.com}, and therefore belong to the same origin according to 
   the same-origin policy.
   \lib{A.html}, shown in Listing~\ref{lst:A}, contains a script and  
   an iframe from \lib{main.com}. The local script \lib{secret.js} contains  sensitive information given  in Listing~\ref{lst:secret}.
  To protect against XSS, the developer \iffull behind \lib{http://main.com} \fi have installed the CSP
   for its main page \lib{A.html}, shown in Listing~\ref{lst:CSPA}.
   
       \begin{lstlisting}[caption={Source code of \lib{http://main.com/A.html}.}, label={lst:A}]
	<html>
	  <script src="secret.js"></script>
	   ...
	  <iframe src="B.html"></iframe> 
	</html>
      \end{lstlisting}

    \begin{lstlisting}[caption={Source code of \lib{secret.js}.}, label={lst:secret}]
	  var secret = "42";
   \end{lstlisting}

   \begin{lstlisting}[caption={CSP of \lib{http://main.com/A.html}.}, label={lst:CSPA}]
Content-Security-Policy: default-src 'none'; 
script-src 'self'; child-src 'self'
\end{lstlisting}
   This CSP provides an effective protection against XSS:

   \subsubsection{Only parent page has CSP}   
   \label{sec:onlyparent}
   
    According to the latest version of CSP\footnote{\url{https://www.w3.org/TR/CSP2/\#which-policy-applies}}, 
   only the CSP of the iframe applies to its content, and it ignores completely the CSP of the including page.
   In our case, if there is no CSP in \lib{B.html} then its resource loading is not restricted.
   As a result, an iframe \lib{B.html} without CSP is potentially vulnerable to XSS, since
   any injected code may be executed within \lib{B.html} with no restrictions. 
   Assume \lib{B.html} was exploited by an attacker injecting a script \lib{injected.js}. 
   Besides taking control over \lib{B.html}, this attack now propagates to the including page \lib{A.html}, as we show in  Fig.~\ref{fig:attack}.
   The XSS attack extends to the including parent page because of the inconsistency 
   between the CSP and SOP.
   When a parent page and an iframe are from the same origin according 
   to SOP, a parent and an iframe share the same privileges and can access each other's code 
   and resources.

 For our example,  \lib{injected.js} is shown in Listing~\ref{lst:injected}. 

 This script executed in \lib{B.html} retrieves the secret value from its parent page 
 (\code{parent.secret}) and transmits it 
 to an attacker's server \lib{{http://attacker.com}} via XMLHttpRequest\footnote{The XMLHttpRequest 
 is not forbidden by the SOP for \lib{B.html} because an attacker has activated the Cross-Origin 
 Resource Sharing mechanism\cite{CORS-W3C} on her server \code{http://attacker.com}.}. 

\begin{lstlisting}[caption={Source code of \lib{injected.js}.}, label={lst:injected}]   
function sendData(obj, url){
	var req = new XMLHttpRequest();
	req.open('POST', url, true);
	req.send(JSON.stringify(obj));
}
sendData({secret: parent.secret}, 'http://attacker.com/send.php'); 
\end{lstlisting}

A straightforward solution to this problem is to ensure that the protection 
mechanism for the parent page also propagates to the iframes 
from the same domain. Technically, it means that the CSP of the iframe
should be the same or more restrictive than the CSP of the parent. 
In the next example we 
show that this requirement does not necessarily prevent possible CSP 
violations due to SOP.

\subsubsection{Only iframe page has CSP}
\label{sec:onlyiframe}

Consider a different web application, where the including parent page \lib{A.html} does not 
have a CSP, while its iframe \lib{B.html} contains a CSP from Listing~\ref{lst:CSPA}.
In this example, \lib{B.html}, shown in Listing~\ref{lst:B} now contains some sensitive 
information stored in \lib{secret.js} (see Listing~\ref{lst:secret}). 

       \begin{lstlisting}[caption={Source code of \lib{http://main.com/B.html}.}, label={lst:B}]
	<html>
	   ...
	  <script src="secret.js"></script> 
	</html>
      \end{lstlisting} 
   
Since the including page \lib{A.html} now has no CSP, it is potentially vulnerable to XSS, 
and therefore may have a malicious script \lib{injected.js}.
The iframe \lib{B.html} has a restrictive CSP, that effectively contributes to protection 
against XSS. 
Since \lib{A.html} and \lib{B.html} are from the same origin, the malicious
injected script can profit from this and steal sensitive information from \lib{B.html}. 
For example, the script may call the \code{sendData} function with the secret information:
\begin{lstlisting}
sendData({secret: children[0].secret}, 'http://attacker.com/send.php'); 
\end{lstlisting}  

Thanks to SOP, the script \lib{injected.js} fetches the secret from it's 
child iframe \lib{B.html} and sends it to \lib{http://attacker.com}. 

\subsubsection{CSP violations due to origin relaxation} 
\label{sec:documentdomain}

A page may change its own origin with some limitations. 
By using the \code{document.domain} API, the script can change 
its current domain to a superdomain. As a result, 
a shorter domain is used for the subsequent origin 
checks\footnote{\url{https://developer.mozilla.org/en-US/docs/Web/Security/Same-origin_policy\#Changing_origin}}.

Consider a slightly modified scenario, where the main page \lib{A.html} from 
\lib{http://main.com} includes an iframe 
\lib{B.html} from its sub-domain \lib{http://sub.main.com}. Any script in \lib{B.html} 
is able to change the origin to \lib{http://main.com} by executing the following line:

\begin{lstlisting}
document.domain = "main.com";
\end{lstlisting}
If \lib{A.com} is willing to communicate with this iframe, it should also execute 
the above-written code so that the communication with \lib{B.html} will be possible.
The content of \lib{B.html} is now treated by the web browser as the same-origin content 
with \lib{A.html}, and therefore any of the previously described attacks become possible.

 \subsubsection{Categories of CSP violations due to SOP}
  \label{sec:categories}
  
  We distinguish three different cases when the CSP violation might occur because of SOP:
\begin{description}
\item[Only parent page or only iframe has CSP] A parent page and an iframe page
are from the same origin, but only one of them contains a CSP. 
The CSP may be violated due to the unrestricted access of a page without CSP 
to the content of the page with CSP. 
We demonstrated this example in Sections~\ref{sec:onlyparent} and \ref{sec:onlyiframe}. 
\item[Parent and iframe have different CSPs] A parent page and an iframe page
are from the same origin, but they have different CSPs. Due to SOP, 
the scripts from one page can interfere with the content of another page thus violating 
the CSP.
\item[CSP violation due to origin relaxation] A parent page 
and an iframe page have the same higher level domain, port and scheme, but however 
they are not from the same origin. Either CSP is absent in one of them, or they have different 
CSPs -- in both cases CSP may be violated because the pages can relax their origin to the 
high level domain by using \code{document.domain} 
API, as we have shown in Section~\ref{sec:documentdomain}. 
\end{description}

\section{Empirical study of CSP violations}\label{sec:study}

We have performed a large-scale study on the top 10,000 Alexa sites  to 
detect whether CSP may be violated due to an inconsistency between CSP and SOP. 
For collecting the data, we have used CasperJS\cite{CasperJS} on top of PhantomJS headless browser\cite{PhantomJS}. 
 The User-Agent HTTP header  was instantiated as a recent Google Chrome browser.

\subsection{Methodology}

  \begin{figure*}[!tp]
    \includegraphics[width=1\textwidth]{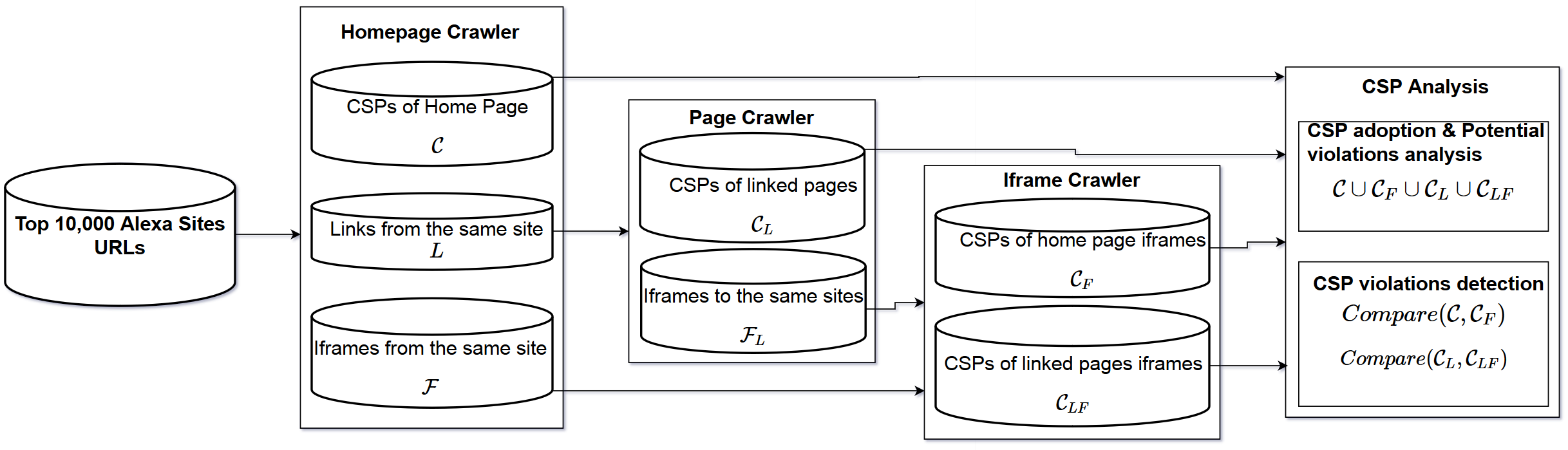}
    \caption{Data Collection and Analysis Process}
    \label{fig:framework}
    \end{figure*}
    
The overview of our data collection and CSP comparison process is given  in Figure~\ref{fig:framework}. The main difference in our data collection process from previous works on CSP measurements in the wild\cite{Weic-etal-16-CCS,Calz-Rabi-Bugl-16-CCS} is that we 
crawl not only the main pages of each site, but also other pages. First, we collect pages accessible through links of the main page and pointing to the same site. Second, to detect possible CSP violations due to SOP, we have collected all the iframes present on the home pages and linked pages.

\subsubsection{Data Collection}

 \iffull  
   We run PhantomJS using as user agent \textit{Mozilla/5.0 (X11; Linux x86\_64) AppleWebKit/537.36 (KHTML, like Gecko) Chrome\/51.0.2704.63 Safari/537.36}. The study was performed on an internal cluster of 200 cores, using OpenMP to benefit from parallelization.
  \fi

\textbf{ Home Page Crawler} 
For each site in top 10,000 Alexa list, we crawl the home page, parse its source code and 
extract three elements: 
\begin{enumerate*}[(1)]
   \item a CSP of the site's home page stored in HTTP header as well as in \code{<meta>} HTML tag;
   we denote the CSPs of the home page by $\CSPH$;
   \item to extract more pages from the same site, we analyse the source of the links 
   via \code{<a href=...>} tag and extract URLs that point to the same site, we denote 
   this list by $\Links$.
   \item we collect URLs of iframes present on the home page via \code{<iframe src=...>} tag and 
   record only those belonging to the same site, we denote this set by $\HomeFrames$. 
\end{enumerate*}

{\bf Page Crawler} 
We crawl all the URLs from the list of pages $\Links$, and for each page we repeat the 
process of extraction of CSP and relevant iframes, similar to the steps (1) and (3) of the home 
page crawler. As a result, we get a set of CSPs of linked pages $\CSPL$ and 
a set of iframes URLs $\LinkedFrames$ that we have extracted from the linked pages in $\Links$.
    
{\bf Iframe Crawler}

  For every iframe URL present in the list of home page iframes $\Frames_H$, and in the list 
  of linked pages iframes $\Frames_L$, we extract their corresponding CSPs and store in 
  two sets: $\CSPFH$ for home page iframes and $\CSPFL$ for linked page iframes. 
 
\subsubsection{CSP adoption analysis} 
  Since CSP is considered an effective countermeasure for a number of web attacks, 
  programmers often use it to mitigate such attacks on the main pages of their sites.
  However, if CSP is not installed on some pages of the same site, this can potentially leak to
  CSP violations due to the inconsistency with SOP when another page from the 
  same origin is included as an iframe (see Figure~\ref{fig:attack}). 
   In our database, for each site, we recorded its home page, a  number of linked pages
  and iframes from the same site. This allows us to analyse how CSP is adopted at every 
  popular site by checking the presence of CSP on every crawled page and iframe of each site. 
  To do so, we analyse the extracted CSPs: $\CSPH$ for the home page, \CSPL\ for linked pages, 
  \CSPFH\ for home page iframes, and \CSPFL\ for linked pages iframes. 
  
\subsubsection{CSP violations detection}

To detect possible CSP violations due to SOP, we have analysed home pages and linked pages 
from the same site, as well as iframes embedded into them. 

{\bf CSP Selection}

To detect CSP violations, we  first remove all the sites where no parent page and 
no iframe page contains a CSP. For the remaining sites, we  pointwise compare
(1) the CSPs of the home pages $\CSPH$ and CSPs of iframes present on these pages \CSPFH; 
(2) the CSPs of the linked pages \CSPL\ and CSPs of their iframes \CSPFL.
To check whether a parent page CSP and an iframe CSP are equivalent, 
we have applied the CSP comparison algorithm (Figure ~\ref{fig:framework})

{\bf CSP Preprocessing}
We first normalise each CSP policy, by splitting it into its directives. 
\begin{itemize}
 \item If \textbf{default-src} directive is present (\textbf{default-src} is a fallback for most of the other directives), 
then we extract the source list $s$ of \textbf{default-src}. We analyse which directives are missing in the CSP, and 
explicitly add them with the source list $s$. 

\item If \textbf{default-src} directive is absent, we extract missing directives from the CSP. In this case, 
there are no restrictions in CSP for every absent directive. We therefore explicitly add them with the 
most permissive source list. A missing \scrsrc is assigned \star \uninline \uneval as the most permissive source list \cite{CSP2-W3C}.
 
 \item In each source list, we modify the special keywords: 
 (i) \self is replaced with the origin of the page containing the CSP;
 (ii) in case of \uninline with hash or nonce, we remove \uninline from the directive since 
 it will be ignored by the CSP2.
 (iii)  \none keywords are removed from all the directives;
 (iv) nonces and hashes are removed from all the directives since they cannot be compared;
(iv) each whitelisted domain is extended with a list of schemes and port numbers from the URL of  
the page includes the CSP\footnote{For example, according to CSP2, if the page scheme is \code{https}, 
and a CSP contains a source \code{example.com}, then the user agent should allow content only 
from \code{https://example.com}, while if the current scheme is \code{http}, it would allow 
both \code{http://example.com} and \code{https://example.com}.}. 
\end{itemize}

 {\bf CSP Comparison} \label{sec:csp_comparison_algo}
We compare all the directives present in the two CSPs to identify whether the two policies 
 require the same restrictions. Whenever the two CSPs are different, our algorithm returns
 the names of directives that do not match. The demonstration of the comparison is accessible
  on\cite{CSP_Violations}. For each directive in the policies we compare the source lists and the algorithm proceeds if the 
elements of the lists are identical in the normalised CSPs.

\subsubsection{Limitations}
Our methodology and results have two(2) limitations that we explain here. 

{\bf User interactions}
The automatic crawling process did not include any real-user-like interactions with top sites. 
As such the set of iframes and links URLs we have analysed is an underestimate of all links and iframes a site may contain.

{\bf Pairs of (parent-iframe)}
In this study, we consider CSP violations in same origin (parent, iframe) couples only. Their are though further combinations such as couples
of sibling iframes in a parent page that we could have considered. 
Overall, our results are conservative, since the problem might have been worst without those limitations.

  \label{sec:stats}
  \subsection{Results on CSP Adoption}

The crawling of Alexa top 10,000 sites was performed in the end of August, 2016. 
To extract several pages from the same site, we have also crawled all the links and iframes on a page that 
point to the same site. In total, we have gathered 1,090,226 from 9,885 different sites. 
On median, from each site we extracted 45 pages, with a maximum number of 9,055 pages  
found on \url{tuberel.com}.
Our crawling statistics is presented in Table \ref{table:statistics}. 
  \begin{table}
  \begin{tabular}{|p{5.5cm}|l|}
  \hline
  Sites successfully crawled & 9,885 \\ \hline
  Pages visited & 1,090,226 \\ \hline
  Pages with iframe(s) from the same site  & 648,324 \\ \hline 
  Pages with same-origin iframe(s) & 92,430 \\ \hline
  Pages with same-origin iframe(s) where page and/or iframe has CSP & 692 \\ \hline

  Pages with CSP & 21,961 (2.00\%) \\ \hline 
  Sites with CSP on home page & 228 (2.3\%) \\ \hline 
  Sites with CSP on some pages & 523 (5.29\%) \\ \hline

  \end{tabular}
  \caption{Crawling statistics}
  \label{table:statistics}
  \SHORTEN
  \end{table}
More than half of the pages contain an iframe, and 13\% of 
pages do contain an iframe from the same site. This indicates the potential 
surface for the CSP violations, when at least one page on the site has a CSP 
installed. We discuss such potential CSP violation in details in Section~\ref{sec:potential}.
Similarly to previous works on CSP adoption\cite{Weic-etal-16-CCS,Calz-Rabi-Bugl-16-CCS}, 
we have found that CSP is present on only 228 out of 9,885 home pages (2.31\%). 
While extending this analysis to almost a million pages, we have found a similar 
rate of CSP adoption (2.00\%). 

    \begin{figure}[!tp]
    \includegraphics[height=3cm, width=0.4\textwidth]{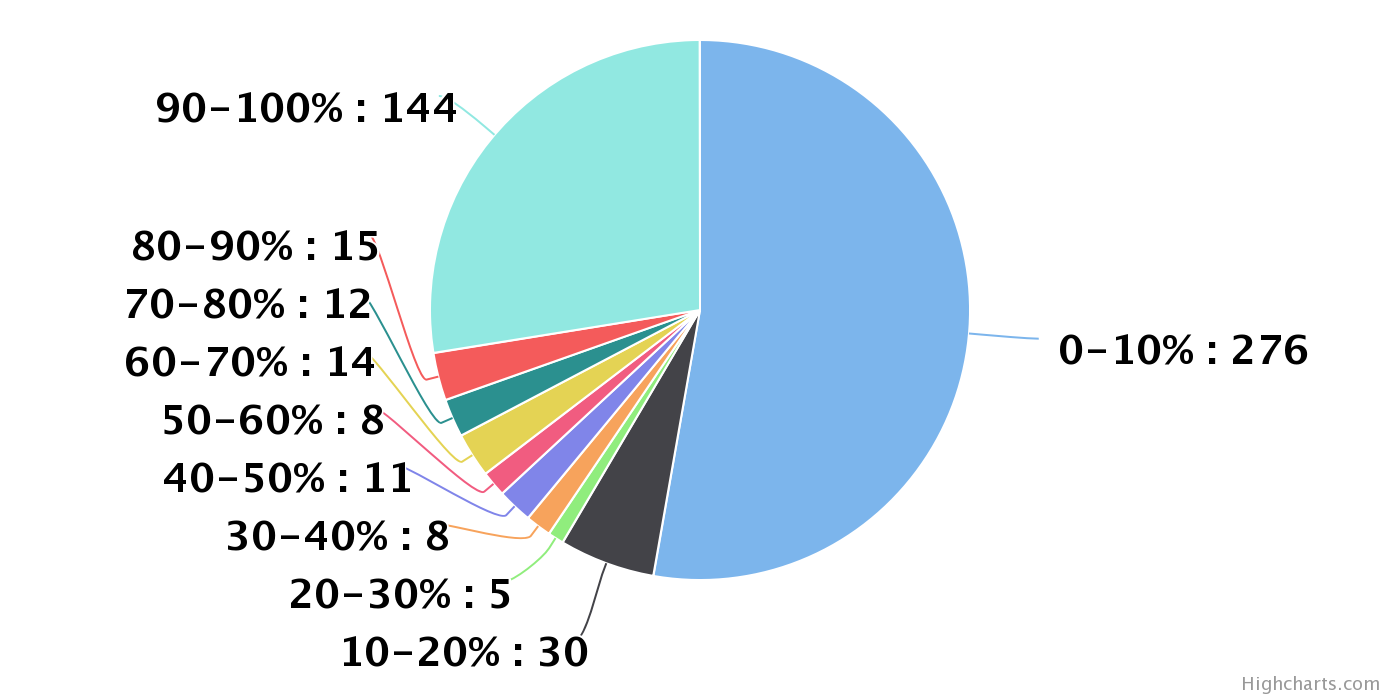}
    \caption{Percentage of pages with CSP per site}
    \label{fig:csp_per_site}
    \SHORTEN
   \end{figure} 
   
Differently from previous studies that anlaysed only home pages, or only pages in separation, 
we have analysed how many sites have at least some pages that adopted CSP. 
We have grouped all pages by sites, and found that 5.29\% of sites contain some pages with CSPs.  
It means that CSP is more known by the website developers, but for some reason is not widely adopted on 
all the pages of the site.

We have then analysed how many pages on each site have adopted CSPs. 
For each of 523 sites, we have counted how many pages (including 
home page, linked pages and iframes) have CSPs. 
Figure~\ref{fig:csp_per_site} shows that more than half of the sites 
have a very low CSP adoption on their pages: on 276 sites out of 529, 
CSP is installed on only 0-10\% of their pages. 
This becomes problematic if other pages without CSP are not XSS-free.
However, it is interesting to note that around a quarter of sites do 
profit from CSP by installing it on 90-100\% of their pages.

\subsection{Results on CSP violations due to SOP}

As described in Section~\ref{sec:categories}, we distinguish several categories 
of CSP violations when a parent page and an iframe on this page are from the 
same origin according to SOP. To account for possible CSP violations, we only consider 
cases when either parent, or iframe, or both have a CSP installed.
From all the 21,961 pages that have CSP installed, we 
have removed the pages, where CSPs are in report-only mode, 
having left 18,035 pages with CSPs in enforcement mode.
    
 \begin{table*}
 \centering
      \begin{tabular}{l|l|l|l|l|l|}
	\cline{2-4} 
	 & \multicolumn{1}{c|}{Same-origin parent-iframe} & \multicolumn{1}{c|}{Possible to relax origin}  & Total\\
	
	\hline
	\multicolumn{1}{ |c|  }{Only parent page has CSP} & {\bf 83} & {\bf 1388} & 1471\\
	\hline
	\multicolumn{1}{ |c|  }{Only iframe has CSP} & {\bf 16}  & {\bf 240}  & 256\\
	\hline
	\multicolumn{1}{ |c|  }{Different CSPs in parent page and iframe} & {\bf 70}  & {\bf 44} & 114 \\
	
	\hline

	\multicolumn{1}{ |c|  }{ No CSP violations} &  551 & 109 & 660 \\
	
	\hline	
	\hline
	\multicolumn{1}{ |c|  }{{\bf CSP violations total}} & {\bf 169 (23.5\%)} & {\bf 1672 (94\%)} & 1841\\
	\hline
	
      \end{tabular}
      \caption{Statistics CSP violations due to Same-Origin Policy}
      \label{table:csp_violations}
      \end{table*}
      
    \begin{table*}
      \begin{tabular}{p{3.2cm}|p{4.7cm}|p{8.2cm}|}
	\cline{2-3} 
	& {Same-origin parent-iframe} & {Possible to relax origin} \\
	
	\hline
	\multicolumn{1}{ |c|  }{Only parent page CSP} & yandex.ru & twitter.com, yandex.ru,  mail.ru \\
	\hline
	\multicolumn{1}{ |c|  }{Only iframe CSP} & amazon.com, imdb.com &  --* \\
	
	\hline
	\multicolumn{1}{ |c|  }{Different CSP} &  twitter.com & --* \\
	\hline	
      \end{tabular}
      \\
      {\small *Not found in top 100 Alexa sites.}
      \caption{Sample of sites with CSP violations due to Same-Origin Policy}
      \label{tab:csp_violations_examples}
      \end{table*}
Table~\ref{table:csp_violations} presents possible CSP violations due to SOP.

We have extracted the parent-iframe couples that might cause 
a CSP violation because either (1) only parent or only iframe installed a CSP, or 
(2) both installed different CSPs. First, to account for direct violations because of SOP, 
we distinguish couples where parent and iframe are from the same origin (columns 
2,3), we have found 720 cases of such couples. Second, we analyse possible 
CSP violations due to origin relaxation: we have collected 1781 
couples that are from different origins but their origins can be relaxed 
by \code{document.domain} API (see more in Section~\ref{sec:documentdomain}) -- 
these results are shown in column 3.

In Table~\ref{tab:csp_violations_examples} we present the names of the 
domains out of top 100 Alexa sites, where we have found different CSP violations. 
Each company in this table have been  notified about the possible CSP violation.
Concrete examples of the page and iframe URLs and 
their corresponding CSPs for each such violation can be found in the
corresponding technical report\cite{CSP-SOP-full}. 
All the collected data is available online\cite{CSP_Violations}.

{\bf CSP violations in presence of \code{document.domain}}
According to our results, in presence of \code{document.domain}, 94\% of (parent, iframe) pages can  have their CSP violated.
Those violations can occur only if both parent and iframes pages execute \code{document.domain} to the same top level domain.
Thus, our result is an over-approximation, assuming that \code{document.domain} is used in all of those pages and iframes.
According to\cite{Chrome-Platform-Status}, \code{document.domain} is used in less than 3\% of web pages.

\subsubsection{Only parent page or only iframe has CSP}
\label{sec:result_oneCSP}

We first consider a scenario when a parent page and an iframe are from the same origin, 
but only one of them contains a CSP. 
Intuitively, if only a parent page has CSP, then an iframe can violate CSP by executing 
any code and accessing the parent page's DOM, inserting content, access cookies etc. 
Among 720 parent-iframe couples from \emph{the same origin}, we have found 83 cases (11.5\%) 
when only parent has a CSP, and 16 cases (2.2\%) when only iframe has a CSP. 
These CSP violations originate from 13 (for parent) and 4 (for iframe) sites. 
For example, such possible violations 
are found on some pages of amazon.com, yandex.ru and imdb.com
(see Table~\ref{tab:csp_violations_examples}). 
CSP of a parent or iframe may also be violated because of \emph{origin relaxation}. 
We have identified 1388 cases (78\%) of parent-iframe couples where such 
violation may occur because CSP is present only in the parent page. 
This was observed on 20 different sites, including twitter.com, yandex.ru 
and others. Finally, in 240 cases (13.5\%) only iframe has CSP installed, which was found 
on 11 different sites. We manually checked the parent and iframes involved in CSP violations for sites in Table~\ref{tab:csp_violations_examples}.
In all of those sites, either the parent or the iframe page is a login page\cite{CSP_Violations}.
We furthermore checked how effective are the CSP of those pages, using CSPEvaluator\footnote{https://csp-evaluator.withgoogle.com/}, proposed by Lukas et al.\cite{Weic-etal-16-CCS}.
and found out that the CSP policies involved in these are moreover all bypassable.

\subsubsection{Parent and iframe have different CSPs}
\label{sec:result_diffCSP}     
    
In a case when a page and iframe are from the same origin, but their 
corresponding CSPs are different, may also cause a violation of CSP.  
From the 720 \emph{same-origin} parent-iframe couples, we have found 
70 cases (9.7\%) (from 3 sites)  when their CSPs differ, and for \emph{an origin relaxation} (from 6 sites)
case, we have identified only 44 such cases (2.5\%).
This setting was found on some pages of twitter.com for instance.

      \begin{figure}[!t]
      \includegraphics[width=0.45\textwidth]{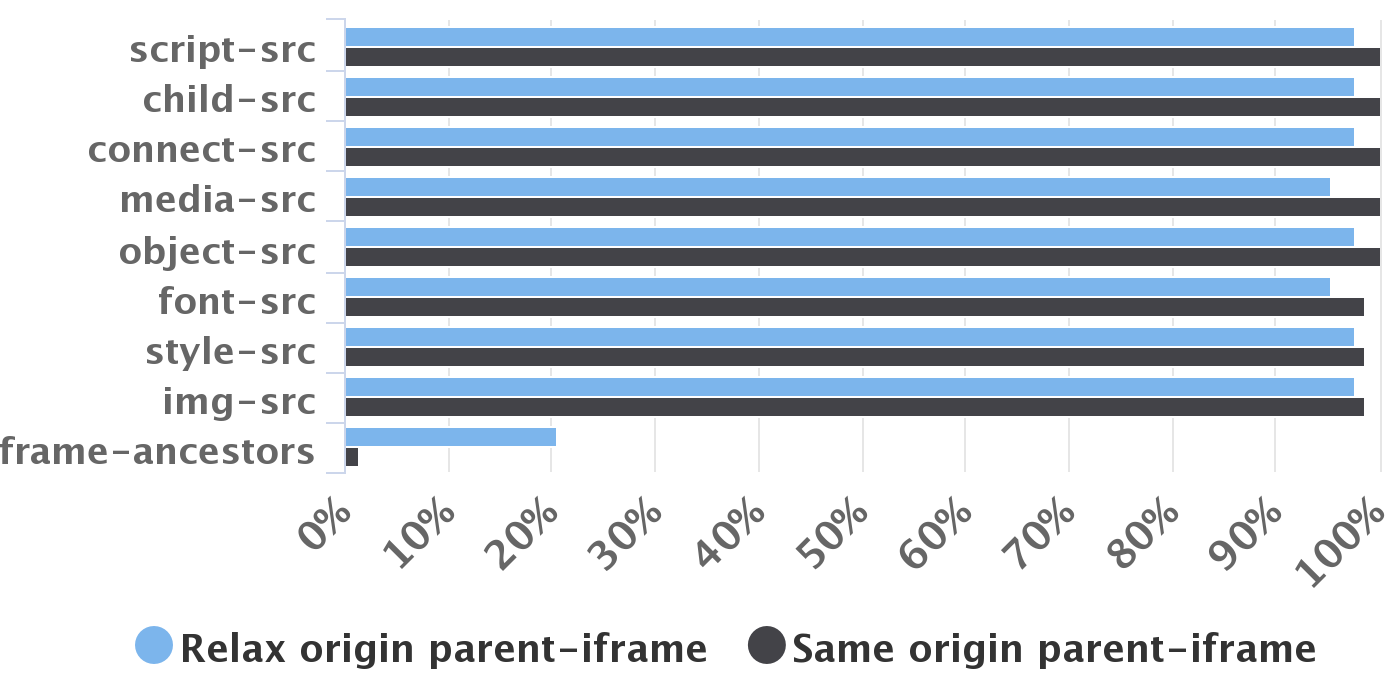}
      \caption{Differences in CSP directives for parent and iframe pages}
      \label{fig:csp_violations_directives_differences}
      \SHORTEN
      \end{figure} 
      
We have further analysed the differences in CSPs found on parent and iframe 
pages. For all the 114 pairs of parent-iframe (either same-origin or 
possible origin relaxation), we have compared CSPs they installed, 
directive-by-directive. Figure \ref{fig:csp_violations_directives_differences}  
shows that every parent CSP and iframe CSP differ on almost every directive 
 -- between 90\% and 100\%. 
The only exception is \frasrc directive, which is almost the same 
in different parent pages and iframes. If properly set, 
this directive gives a strong protection against clickjacking attacks, 
therefore all the pages of the same origin are equally protected.

 \subsubsection{Potential CSP violations} 
 \label{sec:potential}
 
A \emph{potential CSP violation} may happen when in a site, either some pages have CSP and some others do not, or pages have different CSP.
 When those pages get nested as parent-iframe, we can run into CSP violations, just like in the direct CSP violations cases we have just reported above.
 To analyse how often such violations may occur, we have analysed the 18,035 pages that 
have CSP in enforcement mode. These pages originate  from 729 different origins spread over 442 sites. 
        \begin{table*}
   \begin{tabular}{l|l|l|l|}
	\cline{2-4}
	 & Pages & Origins & Sites \\ 
	\hline
	\multicolumn{1}{ |l|  }{A same origin page has no CSP} & 4381 & 197 & 197 \\ \hline
	\multicolumn{1}{ |l|  }{A same origin page has a different CSP} & 1223 & 23 & 23 \\ \hline
	\multicolumn{1}{ |l|  }{\bf Total Potential violations due to same origin pages} & {\bf 5604 (31.1\%)} & {\bf -} & {\bf -} \\ \hline \hline
	\multicolumn{1}{ |l|  }{A same origin (after relaxation) page has no CSP} & 4728 & 340 & 183 \\ \hline
	\multicolumn{1}{ |l|  }{A same origin (after relaxation) has a different CSP} & 2567 & 135 & 44 \\ \hline
	\multicolumn{1}{ |l|  }{\bf Total Potential violations due to same origin (after relaxation} & {\bf 7295(40.4\%)} & {\bf -} & {\bf -} \\ \hline \hline
	\multicolumn{1}{ |l|  }{\bf Potential violations total} & 12899 (72\%)
& 591 (81\%)
	& 379 (52\%)
	\\
	\hline
      \end{tabular}
      \caption{Potential CSP violations in pages with CSP}
      \label{table:potential_csp_violations}
      \vspace*{-0.3cm}
      \end{table*}
Table \ref{table:potential_csp_violations} shows that 72\% of CSPs (12,899 pages)
can be potentially violated, and these CSPs originate from pages of 379 different sites (85.75\%). 
To detect these violations, for each page with a CSP in our database, we have analysed whether there exists 
another page from the same origin, that does not have CSP. 
This page could embed the page with CSP and violate it because of SOP. 
We have detected 4381 such pages (24\%) from 197 origins. 
Similarly, we detected 1223 pages (7\%) when there are same-origin 
pages with a different CSP.
Similarly, we have analysed when potential CSP violations  may happen 
due to origin relaxation. We have detected 4728 pages (26\%), whose CSP may 
be violated because of other pages with no CSP, and 2567 pages (14\%), whose 
CSP may be violated because of different CSP on other relaxed-origin pages. 

  \begin{figure}[!t]
      \includegraphics[width=0.45\textwidth, keepaspectratio]{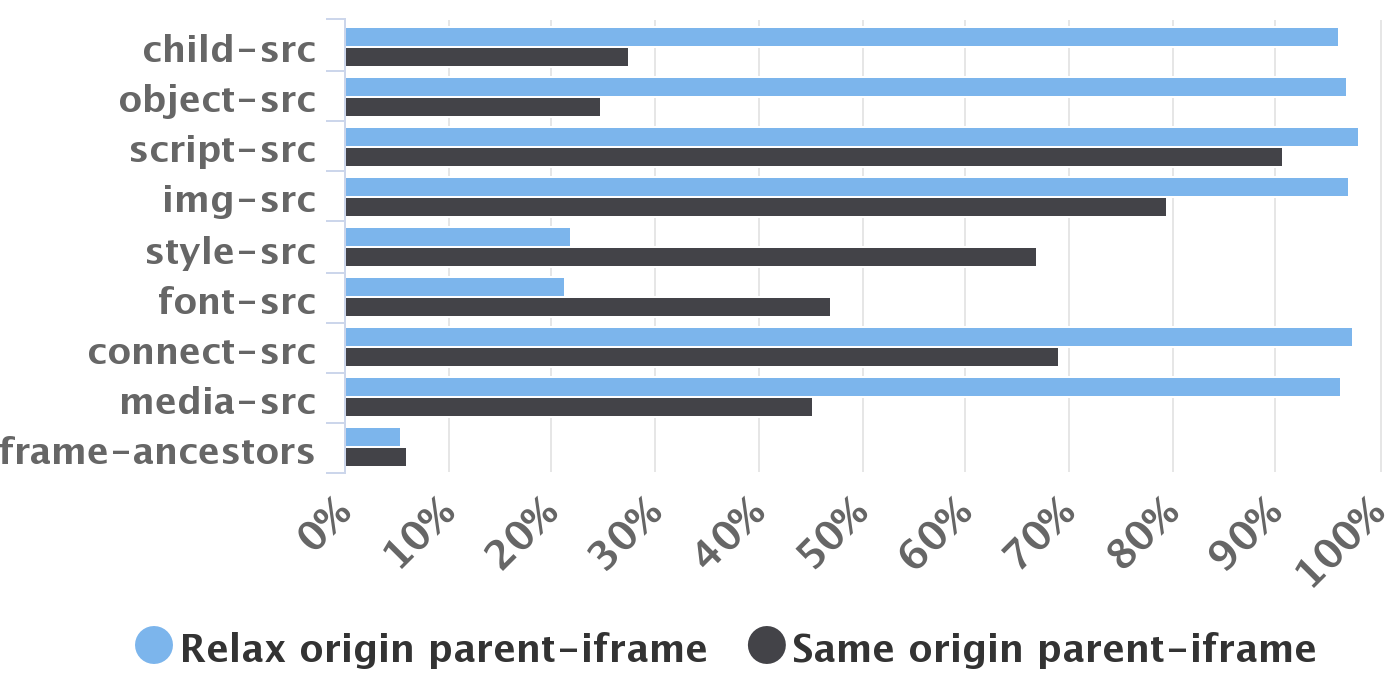}
      \caption{Differences in CSP directives for same-origin and relaxed origin pages}
      \label{fig:same_origin_pages_directives_differences}
      \SHORTEN
   \end{figure} 
 For the pages that have different CSPs, we have compared how much CSPs differ. 
 Figure~\ref{fig:same_origin_pages_directives_differences} shows that CSPs mostly 
 differ in \code{script-src} directive, which protects pages from XSS attacks. This means, 
 that if one page in the origin does whitelist an attacker's domain or an insecure endpoints \cite{Weic-etal-16-CCS}, all the other pages 
 in the same origin become vulnerable because they may be inserted as an iframe to 
 the vulnerable page and their CSPs can be easily violated. 
 
\subsection{Responses of websites owners}
We have reported those issues to a sample of sites owners, using either HackerOne\footnote{https://hackerone.com}, or contact forms when available.
Here are some selected quotes from our discussions with them.
 \begin{quote}
  ``\textit{Yes, of course we understand the risk that under some circumstances XSS on one domain can be used to bypass CSP on another domain, but it's simply impossible to implement CSP across all (few hundreds) domains at once on the same level. 
  We are implementing strongest CSP currently possible for different pages on different domains and keep going with this process to protect all pages, after that we will strengthen the CSP. We believe it's better to have stronger CSP policy where possible rather than have same weak CSP on all pages or not having CSP at all. Having in mind there are hundreds of domains within mail.ru, at least few years are required before all pages on all domains can have strong CSP.}''  -- \code{Mail.ru}
 \end{quote}

 \begin{quote}
  ``\textit{[...]the sandbox is a defense in depth mitigation[...]. We definitely don't allow relaxing document.domain on www.dropbox.com[...]}'' -- \code{Dropbox.com}
 \end{quote}
 
 \begin{quote}
  ``\textit{While this is an interesting area of research, are you able to demonstrate that this behavior is currently exploitable on Twitter? It appears that the behavior you have described can increase the severity of other vulnerabilities but does not pose a security risk by itself. Is our understanding correct?
  [...]We consider this to be more of a defensive in depth and will take into account with our continual effort to improve our CSP policy}'' -- \code{Twitter.com}
 \end{quote}
   
  \begin{quote}
   ``\textit{I believe we understand the risk as you've described it.}'' -- \code{Imdb.com} 
  \end{quote}

  \label{sec:countermeasures}
    \section{Avoiding CSP Violations}

Preventing CSP violations due to SOP can be achieved by having the \textbf{same} effective CSP for \textbf{all} same-origin pages in a site, and prevent origin relaxation.

{\bf Origin-wide CSP}:
Using CSP for all same-origin pages can be manually done but this solution is error-prone.
A more effective solution is the use of a specification such as Origin Policy\cite{Origin-Policy} in order to set a header for the whole origin.

{\bf Preventing Origin Relaxation}: 
Having an origin-wide CSP is not enough to prevent CSP violations. 
By using origin relaxation, pages from different origins can bypass the SOP\cite{Singh-etal-10-SP}.
Many authors provide guidelines on how to design an effective CSP\cite{Weic-etal-16-CCS}. 
Nonetheless, even with an effective CSP, an embedded page from a different origin in the same site can use \code{document.domain} to relax its origin. 
Preventing origin relaxation is trickier. 

Programmatically, one could prevent other scripts from modifying \code{document.domain} by making a script run first in a page\cite{Swam-etal-14-ACM}. The first script that runs on the page would be:
\begin{lstlisting}
Object.defineProperty(document, "domain", { __proto__: null, writable: false, configurable: false});
\end{lstlisting}
 
A parent page can also indirectly disable origin relaxation in iframes by sandboxing them. This can be achieved by using \sansrc as an attribute for iframes or as directive for the parent page CSP.
Unfortunately, an iframe cannot indirectly disable origin relaxation in the page that embeds it. However, the \frasrc directive of CSP gives an iframe control over the hosts that can embed it. 
Finally, a more robust solution is the use of a policy to deprecate \code{document.domain} as proposed in the draft of Feature policy\cite{Feature-Policy}.
The feature policy defines a mechanism that allows developers to selectively enable and disable the use of various browser features and APIs. 

{\bf Iframe sandboxing}: 
Combining attribute \allowsrc and \allowsop as values for \sansrc successfully disables \code{document.domain} in an iframe\footnote{We found out that \url{dropbox.com} actually puts \sansrc attribute for all its iframes, and therefore avoids the possible CSP violations.
We have had a very interesting discussion on \url{Hackerone.com} with Devdatta Akhawe, a Security Engineer at Dropbox, who told us more about their security practices regarding CSP in particular.}.
We recommend the use of \sansrc as a CSP directive, instead of an HTML iframe attribute. The first reason is that \sansrc as a CSP directive, automatically applies to all iframes that are in a  page, avoiding the need to manually modify all HTML iframe tags. 
Second, the \sansrc directive is not programmatically accessible to potentially malicious scripts in the page, as is the case for the \sansrc attribute (which can be removed from an iframe  programmatically, replacing the sandboxed iframe with another identical iframe but without the \sansrc attribute).

{\bf Limitations}
An origin-wide CSP (the same CSP for all same origin pages) can become very liberal if all same origin pages do not require the same restrictions. 
In order to implement the solution we propose, one needs to consider the intended relation between a parent page and an iframe page, in presence of CSP.
In the case where the two(2) pages should be allowed direct access to each other content, then, since same origin pages can bypass page-specific security characteristics \cite{Jack-Bart-08-W2SP}, the solution is to have the same CSP for both the page and the iframe.
However, if direct access to each other content is not a required feature, one can keep different CSPs in parent and iframe, or have no CSP at all in one of the parties, but their contents should be isolated from each other. The solution here is to use sandboxing.
Nonetheless, there are other means (such as \code{postMessage}) by which one can securely achieve communication between the pages.

\section{Inconsistent Implementations}\label{sec:inconsistencies}

	Combining origin-wide CSP with \allowsrc \textbf{sandbox} directive would have been sufficient at preventing the inconsistencies between CSP and the same origin policy. 
	Unfortunately, we have discovered that for some browsers, this solution is not sufficient.
	Starting from HTML5, major browsers, apart from Internet Explorer, supports the new \srcdoc attribute for iframes. Instead of providing a URL which content will be loaded in an iframe, one provides directly the HTML content of the iframe in the \srcdoc attribute.
	According to CSP2 \cite{CSP2-W3C}, \S 5.2, the CSP of a page should apply to an iframe which content is supplied in a \srcdoc attribute. This is actually the case for all majors browsers, which support the \srcdoc attribute.
	However, there is a problem when the \textbf{sandbox} attribute is set to an \srcdoc iframe.
	 
	\textbf{Webkit}-based\footnote{https://en.wikipedia.org/wiki/WebKit} and \textbf{Blink}-based\footnote{https://en.wikipedia.org/wiki/Blink\_(web\_engine)} browsers (Chrome, Chromium, Opera) always comply with CSP. 
	The CSP of a page will apply to all \srcdoc iframes, even in those iframes which have a different origin than that of the page, because they are sandboxed without \allowsop.
	
	In contrast, we noticed  that  in Gecko-based  browsers (Mozilla Firefox), the CSP of the page applies to that of the srcdoc iframe if and only if allow-same-origin is present as value for the attribute. Otherwise it does not apply. 
	The problem with this choice is the following. A third party script, whitelisted by the CSP of the page, can create a \srcdoc iframe, sandboxing it with \allowsrc only, and load any resource that would normally be blocked by the CSP of the page if applied in this iframe. 
	This way, the third party script successfully bypasses the restrictions of the CSP of the page. Even though loading additional scripts is considered harmless in the upcoming version 3 \cite{CSP3-W3C, Weic-etal-16-CCS} of CSP, this specification says nothing about violations that could 
	occur due to the loading of other resources inside a \srcdoc sandboxed iframe, like resources whitelisted by \objsrc directive for instance, additional iframes etc.
	
	We have notified the W3C, and the Mozilla Security Group. Daniel Veditz, a lead at Mozilla Security Group, recognizes  this as a bug and explains:
	\begin{quote}
	 ``\textit{Our internal model only inherits CSP into same-origin frames (because in theory you’re otherwise leaking info across origin boundaries) and iframe sandbox creates a unique origin. Obviously we need to make an exception here (I think we manage to do the same thing for src=data: sandboxed frames).}''
	\end{quote}

{\bf CSP specification and srcdoc iframes} The problem of imposing a CSP to an unknown page is illustrated by the following example\cite{Embedded-Policy}. If a trusted third party library, whitelisted by the CSP of the page, uses  security libraries inside an isolated context (by sandboxing them in a \srcdoc iframe, setting \allowsrc as sole value for the \textbf{sandbox}) then, the page's CSP will block the security libraries and possibly introduce new vulnerabilities. 
	Because of this, it was unclear to us the intent of CSP designers regarding srcdoc iframes.
	Mike West, one of the CSP editors at the W3C and also Developer Advocate in Google Chrome’s team, clarified this to us:
	
      \begin{quote}
	``\textit{I think your objection rests on the notion of the same-origin policy preventing the top-level document from reaching into it's sandboxed child.
	That seems accurate, but it neglects the bigger picture: \srcdoc documents are produced entirely from the top-level document context.
	Since those kinds of documents are not delivered over the network,
	they don't have the opportunity to deliver headers which might configure their settings.
	We impose the parent's policy in these cases, because for all intents and purposes, the \srcdoc document \emph{is} the parent document.}''
      \end{quote}


  \section{Related Work}

CSP has been proposed by Stamm et al.\cite{Stam-Ster-Mark-10-WWW} as a refinement of SOP\cite{Same-Origin-Policy}, in order to help mitigate Cross-Site-Scripting\cite{Yuso-Path-2016-IEEE} and data exfiltration attacks.
The second version\cite{CSP2-W3C} of the specification is supported by all major browsers, and the third version~\cite{CSP3-W3C}
is under active development.  Even though CSP is well supported~\cite{Calz-Rabi-Bugl-16-CCS}, its endorsement by web sites is rather slow. Weissbacher et al.\cite{Weis-Laui-Robe-14-RAID} performed the first
large scale study of CSP deployment in top Alexa sites, and found that around 1\% of sites were using CSP at the time. A more recent study by Calzavara et al.\cite{Calz-Rabi-Bugl-16-CCS}, show that nearly 8\% of Alexa
top sites now have CSP deployed in their front pages. Another recent study, by Weichselbaum et al.\cite{Weic-etal-16-CCS} come with similar results to the study of Weissbacher et al.\cite{Weis-Laui-Robe-14-RAID}. Our work extends previous results by analysing the adoption of CSP by site not only considering front pages but all the pages in  a site. 
Almost all authors agree that 
CSP adoption is not a straightforward task, and lots of (manual) effort are needed in order to reorganize and modify web pages to support CSP.

Therefore, in order to help web sites developers in adopting CSP, Javed proposed CSP Aider, \cite{Jave-12-W2SP} that automatically crawl a set of pages from a site and propose a site-wide CSP. 
Patil and Frederik\cite{Pati-Fred-16-IJNS} proposed UserCSP, a framework that monitors the browser internal events in order to automatically infer a CSP for a web page based on the loaded resources.
Pan et al.\cite{Pan-etal-16-CCS} propose CSPAutoGen, to enforce CSP in real-time on web pages, by rewriting them on the fly client-side. 
Weissbacher et al.\cite{Weis-Laui-Robe-14-RAID} have evaluated the feasibility of using CSP in report-only mode in order to generate a CSP based on reported violations, or semi-automatically inferring a CSP policy based on the resources that are loaded in web pages. 
They concluded that automatically generating a CSP is ineffective. 
A difficulty which remains is the use of inline scripts in many pages. The first solution is to externalize inline scripts, as can be done by systems like deDacota\cite{DOUP-etal-13-CCS}. Kerschbaumer et al.\cite{Kers-Stam-Brun-16-ICISSP} find that too many pages
are still using \uninline in their CSPs. They propose a system to automatically identify legitimate inline scripts in a page, thereby whitelisting them in the CSP of the underlying page, using script hashes. 
\par Another direction of research on CSP, has been evaluating its effectiveness at successfully preventing content injection attacks. 
 Calzavara et al.\cite{Calz-Rabi-Bugl-16-CCS} found out that many CSP policies in real web sites have errors including typos, ill-formed or harsh policies. Even when the policies are well formed, they have found that
 almost all currently deployed CSP policies are bypassable because of a misunderstanding of the CSP language itself. Patil and Frederik found similar errors in their study\cite{Pati-Fred-16-IJNS}. 
 Hausknecht et al.\cite{Haus-Maga-Sabe-2015-DIMVA} found that some browser extensions, modified the CSP policy headers, in order to whitelist more resources and origins.
 Van Acker et al.\cite{Vana-Haus-Sabe-16-CCS} have shown that CSP fails at preventing data exfiltration specially when resources are prefetched, 
 or in presence of a CSP policy in the HTML meta tag, because the order in which resources are loaded in a web application is hard to predict. Johns\cite{John-14-JISA} proposed hashes for static scripts, and PreparedJS, an extension for CSP, in order to securely handle server-side dynamically generated scripts based on user input.
  Weichselbaum et al.\cite{Weic-etal-16-CCS} have extended nonces and hashes, introduced in CSP level 2\cite{CSP2-W3C}, to remote scripts URLs, specially to tackle the high prevalence of insecure hosts in current CSP policies. 
  Furthermore, they have introduced \textbf{strict-dynamic}. This new keyword states that any additional script loaded by a whitelisted remote script URL is considered a trusted script as well. They also provide guidelines on how to build an effective CSP.  
  Jackson and Barth\cite{Jack-Bart-08-W2SP} have shown that same origin pages can bypass page-specific policies, like CSP. Though, their work predates CSP. 
  To the best of our knowledge, we are the first to explore the interactions between CSP and SOP and report possible CSP violations.

  \section{Conclusions}\label{sec:conclusions}
In this work, we have revealed a new problem that can lead to violations of CSP. We have 
performed an in-depth analysis of the inconsistency that arises due to CSP and  SOP 
and identified three cases when \emph{CSP may be violated}. 

To evaluate how often such violations happen, we performed a large-scale analysis of more 
than 1 million pages from 10,000 Alexa top sites.
We have found that 5.29\% of sites contain pages with CSPs (as opposed to 2\% of home 
pages in previous studies). 

We have also found out that 72\% of current web pages with CSP, are potentially vulnerable to CSP violations. This concerns 379 (72.46\%) sites that deploy CSP.
Further analysing the contexts in which those web pages are used, our results show that when a parent page includes an iframe from the same origin 
according to SOP, in 23.5\% of cases their CSPs may be violated. 
And in the cases where \code{document.domain} is required in both parent and iframes, we identified that such violations may occur in 94\% of the cases.

We discussed measures to avoid CSP violations in web applications 
by installing an origin-wide CSP and using sandboxed iframes.
Finally, our study reveals an inconsistency in browsers implementation of CSP for \code{srcdoc} iframes, that appeared to be a bug in Mozilla Firefox browsers.

\section*{Acknowledgments}
The authors would like to thank the WebAppSec W3C Working Group for useful pointers to related resources at the early stage of this work,
Mike West for very insightful discussions that considerably helped improve this work, Devdatta Akhawe for discussing some security practices at \emph{Dropbox}, and anonymous reviewers and Stefano Calzavara for their valuable comments and suggestions.

 \newpage	
\bibliographystyle{abbrv}
\bibliography{main.bib}

\end{document}